\documentclass[twoside,twocolumn,9pt]{article}
\usepackage{extsizes}
\usepackage[super,sort&compress,comma]{natbib} 
\usepackage[version=3]{mhchem}
\usepackage[left=1.5cm, right=1.5cm, top=1.785cm, bottom=2.0cm]{geometry}
\usepackage{balance}
\usepackage{mathptmx}
\usepackage{sectsty}
\usepackage{graphicx} 
\usepackage{lastpage}
\usepackage[format=plain,justification=justified,singlelinecheck=false,font={stretch=1.125,small,sf},labelfont=bf,labelsep=space]{caption}
\usepackage{float}
\usepackage{fancyhdr}
\usepackage{fnpos}
\usepackage[english]{babel}
\addto{\captionsenglish}{%
  
}
\usepackage{array}
\usepackage{droidsans}
\usepackage{charter}
\usepackage[T1]{fontenc}
\usepackage[usenames,dvipsnames]{xcolor}
\usepackage{setspace}
\usepackage[compact]{titlesec}
\usepackage{hyperref}

\usepackage{epstopdf}

\definecolor{cream}{RGB}{222,217,201}

\begin{document}

\pagestyle{fancy}
\thispagestyle{plain}
\fancypagestyle{plain}{
\renewcommand{\headrulewidth}{0pt}
}

\makeFNbottom
\makeatletter
\renewcommand\LARGE{\@setfontsize\LARGE{15pt}{17}}
\renewcommand\Large{\@setfontsize\Large{12pt}{14}}
\renewcommand\large{\@setfontsize\large{10pt}{12}}
\renewcommand\footnotesize{\@setfontsize\footnotesize{7pt}{10}}
\makeatother

\renewcommand{\thefootnote}{\fnsymbol{footnote}}
\renewcommand\footnoterule{\vspace*{1pt}%
\color{cream}\hrule width 3.5in height 0.4pt \color{black}\vspace*{5pt}} 
\setcounter{secnumdepth}{5}

\makeatletter 
\renewcommand\@biblabel[1]{#1}            
\renewcommand\@makefntext[1]%
{\noindent\makebox[0pt][r]{\@thefnmark\,}#1}
\makeatother 
\renewcommand{\figurename}{\small{Fig.}~}
\sectionfont{\sffamily\Large}
\subsectionfont{\normalsize}
\subsubsectionfont{\bf}
\setstretch{1.125} 
\setlength{\skip\footins}{0.8cm}
\setlength{\footnotesep}{0.25cm}
\setlength{\jot}{10pt}
\titlespacing*{\section}{0pt}{4pt}{4pt}
\titlespacing*{\subsection}{0pt}{15pt}{1pt}

\fancyfoot{}
\fancyfoot[LO,RE]{\vspace{-7.1pt}\includegraphics[height=9pt]{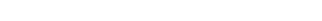}}
\fancyfoot[CO]{\vspace{-7.1pt}\hspace{13.2cm}\includegraphics{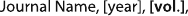}}
\fancyfoot[CE]{\vspace{-7.2pt}\hspace{-14.2cm}\includegraphics{head_foot/RF}}
\fancyfoot[RO]{\footnotesize{\sffamily{1--\pageref{LastPage} ~\textbar  \hspace{2pt}\thepage}}}
\fancyfoot[LE]{\footnotesize{\sffamily{\thepage~\textbar\hspace{3.45cm} 1--\pageref{LastPage}}}}
\fancyhead{}
\renewcommand{\headrulewidth}{0pt} 
\renewcommand{\footrulewidth}{0pt}
\setlength{\arrayrulewidth}{1pt}
\setlength{\columnsep}{6.5mm}
\setlength\bibsep{1pt}

\makeatletter 
\newlength{\figrulesep} 
\setlength{\figrulesep}{0.5\textfloatsep} 

\newcommand{\topfigrule}{\vspace*{-1pt}%
\noindent{\color{cream}\rule[-\figrulesep]{\columnwidth}{1.5pt}} }

\newcommand{\botfigrule}{\vspace*{-2pt}%
\noindent{\color{cream}\rule[\figrulesep]{\columnwidth}{1.5pt}} }

\newcommand{\dblfigrule}{\vspace*{-1pt}%
\noindent{\color{cream}\rule[-\figrulesep]{\textwidth}{1.5pt}} }

\makeatother

\twocolumn[
  \begin{@twocolumnfalse}
{\includegraphics[height=30pt]{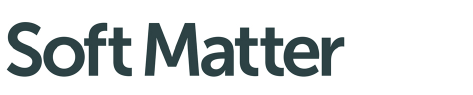}\hfill\raisebox{0pt}[0pt][0pt]{\includegraphics[height=55pt]{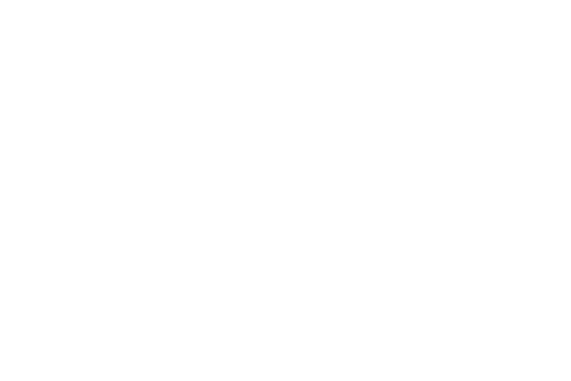}}\\[1ex]
\includegraphics[width=18.5cm]{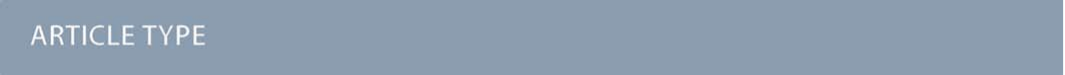}}\par
\vspace{1em}
\sffamily
\begin{tabular}{m{4.5cm} p{13.5cm} }

\includegraphics{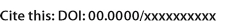} & \noindent\LARGE{\textbf{Collective Dynamics of Intelligent Active Brownian Particles with Visual Perception and Velocity Alignment in 3D: Spheres, Rods, and Worms$^\dag$}} \\
\vspace{0.3cm} & \vspace{0.3cm} \\

 & \noindent\large{Zhaoxuan Liu and Marjolein Dijkstra$^{\ast}$} \\

\includegraphics{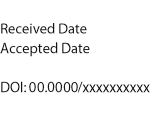} & \noindent\normalsize{Many living systems, such as birds and fish, exhibit collective behaviors like flocking and swarming. Recently, an experimental system of active colloidal particles has been developed, where the motility of each particle is adjusted based on its visual detection of surrounding particles. These particles with visual-perception–dependent motility exhibit group formation and cohesion. Inspired by these behaviors, we investigate intelligent active Brownian particles (iABPs) equipped with visual perception and velocity alignment in three dimensions using computer simulations. The visual-perception-based self-steering describes the tendency of iABPs to move toward the center of mass of particles within their visual cones, while velocity alignment encourages alignment with neighboring particles. We examine how the behavior varies with the visual cone angle $\theta$, self-propulsion speed (Péclet number ${\textrm{Pe}}$), and the interaction strengths of velocity alignment ($\Omega_a$) and visual-based self-steering ($\Omega_v$). Our findings show that spherical iABPs form clusters, worm-like clusters, milling behaviors, and dilute-gas phases, consistent with 2D studies. By reducing the simulation box size, we observe additional structures like band-like clusters and dense baitball formations. Additionally, rod-like iABPs form band-like, worm-like, radiating, and helical structures, while iABP worms exhibit streamlined and blob-like structures. Many of these patterns resemble collective behaviors in nature, such as ant milling, fish baitballs, and worm clusters. Advances in synthetic techniques could enable nanorobots with similar capabilities, offering insights into multicellular systems through active matter.
} \\

\end{tabular}

 \end{@twocolumnfalse} \vspace{0.6cm}

]

\renewcommand*\rmdefault{bch}\normalfont\upshape
\rmfamily
\section*{}
\vspace{-1cm}


\footnotetext{\textit{Soft Condensed Matter \& Biophysics group, Debye Institute for Nanomaterials Science, Utrecht University, Princetonplein 1, 3584 CC Utrecht, the Netherlands. E-mail: m.dijkstra@uu.nl.}}

\footnotetext{\dag~Supplementary Information available: [details of any supplementary information available should be included here]. See DOI: 10.1039/cXsm00000x/}


\section{Introduction}
Active matter consists of entities that extract energy from their environment and convert it into directed motion. Nature offers many examples of active matter systems,  from bird flocks and buffalo herds to bacteria, chasing white blood cells,  cancerous tumor growth, and embryogenesis.  Many living systems exhibit collective behaviors that resemble flocks or swarms \cite{camazine2020self}. These groups can display unique patterns, like ant milling \cite{schneirla1944unique}, where ants continuously circle around a specific point until they die from exhaustion, a phenomenon referred to  as the 'circle of death'. Reindeer can form cyclones, penguin huddles can crystallize and undergo jamming-unjamming transitions \cite{zitterbart2011coordinated}, and sheep display intricate flocking patterns  \cite{gomez2022intermittent}. While many of these behaviors occur in two-dimensional environments, such as on land or ice, aquatic and aerial settings offer opportunities for three-dimensional behaviors. For instance, fish form  baitballs \cite{lopez2012behavioural}---dense spherical aggregates---as a defense mechanism against predators. Fire ants build bridges and rafts  to navigate their surroundings\cite{tennenbaum2016mechanics,foster2014fire}, and blackworms exhibit rapid transitions from tangled blobs to untangled states for survival \cite{patil2023ultrafast,deblais2023worm}. These collective behaviors are crucial for activities like finding food and avoiding predators, emerging naturally from individual responses to others in the group without centralized control. 

The study of such phenomena offers valuable insights not only into biological processes but also into the design of artificial systems that mimic these behaviors. Over the past two decades,  a wide variety of self-propelled colloidal particles has been synthesized, including magnetic-bead-based colloids that mimic artificial flagella \cite{dreyfus2005microscopic}, catalytic Janus particles \cite{howse2007self,erbe2008various,palacci2010sedimentation,baraban2012transport}, laser-heated metal-capped particles \cite{volpe2011microswimmers}, light-activated catalytic colloidal surfers \cite{palacci2013living}, and platinum-loaded stomatocytes \cite{wilson2012autonomous}, as well as robotic systems spanning scales from micrometers to centimeters. Inspired by nature,  researchers have developed models to simulate collective dynamics, where emerging patterns rely on physical interactions and nonreciprocal information exchanges, such as visual-perception-based self-steering  and velocity alignment. These interactions and communication processes are key to the formation of swarms and flocks. Motion alignment and group cohesion are fundamental characteristics of these collective behaviors, and various models have been proposed to understand how these processes give rise to the intricate structures observed in nature.

One of the simplest models that describe collective motion in systems of self-propelled particles is the Vicsek model proposed in 1995 \cite{vicsek1995novel}. In this model, particles move at a constant speed and update their velocity by aligning with the average direction of neighbouring particles within a certain interaction range, while accounting for some noise.
There are no interactions between  particles in this model, and   the large-scale collective behavior is purely driven by the velocity alignment among particles.  The Vicsek model exhibits a phase transition from a disordered state, where particles move in random directions, to an ordered phase with polar order, where particles move in a coherent direction, upon increasing  the particle density or reducing the  noise. 

The Boids model \cite{reynolds1987flocks}, introduced by Reynolds in 1987, describes the behavior of birds, also called boids, by incorporating three fundamental interaction rules: cohesion, separation, and alignment. These rules indicate that individuals strive to maintain proximity to their peers (cohesion), align their movement direction with the group (alignment), and avoid collisions with nearby individuals (separation).  Additionally, the behavioural zonal model \cite{couzin2002collective,couzin2005effective}, introduced by Couzin in 2002, considers distinct interactions between individuals within three non-overlapping zones: repulsion at close distances, alignment with neighbors, and attraction toward others. This model exhibits collective behaviors such as swarming, milling, and groups with highly aligned motion. 

It is important to highlight that the intricate patterns and structures arise  from simple non-reciprocal interactions between individuals, meaning they violate Newton's third law of  action-reaction symmetry. Remarkably, these complex behaviors emerge without any need for external control or coordination. 

Another class of models involves active Brownian particles (ABPs), which mimick the behavior of self-propelled colloidal particles through  self-propulsion forces and short-range repulsive excluded-volume interactions \cite{mandal2019motility}. In these systems, shape of the particles plays a crucial role in determining the collective behavior. Spherical ABPs often exhibit motility-induced phase separation (MIPS), where uniformly distributed particles undergo phase coexistence between a dense cluster and a dilute phase at sufficiently high self-propulsion speeds \cite{cates2015motility,klamser2018thermodynamic,digregorio2018full,paliwal2020role}. In contrast, elongated ABPs exhibit  distinct non-equilibrium behaviors, such as the formation of  immobile dimers \cite{vutukuri2016dynamic}, motile clusters, swarms \cite{bar2020self,abkenar2013collective,peruani2006nonequilibrium}, and the suppression of motility-induced phase separation due to interparticle torques \cite{van2019interparticle},  emphasizing the impact of particle shape on the collective dynamics of active matter systems.

In 2016, visual perception-based self-steering capabilities were introduced to the active Brownian particle  model \cite{barberis2016large,negi2022emergent}, allowing each particle  to reorient its self-propulsion direction based on other particles within a predefined visual cone. These particles, referred to as intelligent active Brownian particles (iABPs), have since evolved with the integration of the Vicsek model \cite{vicsek1995novel}, enabling iABPs to align their self-propulsion direction with that of their neighbors \cite{negi2024collective}. As a result, these iABPs possess both visual perception-based self-steering  and velocity alignment capabilities. It is also important to mention that these models can be realized experimentally  \cite{bauerle2020formation,bauerle2018self,lavergne2019group}. Recent studies on iABPs  \cite{negi2024collective} have demonstrated a range of  collective structures, including milling, clusters, worms, and dilute-gas formations. However, these investigations have primarily been restricted to two-dimensional spaces with disk-like particles. Given that many animals navigate in three-dimensional environments and often have  non-spherical bodies, there is a compelling case for expanding the study of iABPs to three-dimensional systems and aspherical shapes.

In this paper, we investigate three-dimensional intelligent active Brownian particles (iABPs) using computer simulations to explore their collective behaviors. We analyze these systems as a function of the vision angle $\theta$, which affects the opening angle of the visual cone, the Péclet number ${\textrm Pe}$, which  influences the self-propulsion speed, and the ratio $\Omega_a/\Omega_v$, which balances the velocity alignment and vision-based self-steering capabilities. Our results indicate that spherical iABPs form clusters, milling patterns, worm-like structures, and dilute-gas configurations, consistent with findings in 2D. By reducing the simulation box size, we also observe the emergence of band formations, dispersed clusters, and baitball structures. Additionally, rod-like iABPs exhibit band formations, worm-like, radiating, and helical structures, while iABP worms display band-like, streamlined, and blob-like  configurations.

Some of these structures closely resemble real animal behaviors, such as ant milling, fish baitballs, and worm blobs, indicating that these models have the potential to predict collective animal behaviors. Furthermore, advancements in synthetic techniques could inspire the design of nanorobots and colloidal particles with similar capabilities. Recent research \cite{armengol2023epithelia,forget2022heterogeneous}  suggests that multicellular systems can also be understood through the framework of active matter. If the single-cell organisms that gave rise to multicellular creatures could sense light and velocity, the structures obtained in our model may provide valuable insights into the formation of  multicellular systems.

This paper is organized as follows. In Section~\ref{sec:spheres}, we start our investigation with the collective dynamics of intelligent active Brownian particles (iABPs)  with spherical shapes. In Section~\ref{sec:rods}, we  extend the iABP model to incorporate rod-like shapes using the Kihara potential. Finally,  in Section~\ref{sec:worms}, we  connect spherical iABPs with ordinary Brownian particles to create iABP worms. We end with some conclusions in Section~\ref{sec:conclusions}.

\section{Intelligent active Brownian spheres} 
\label{sec:spheres}

\subsection{Model system}

\begin{figure}[htbp]
    \centering
    \includegraphics[width=8.4cm]{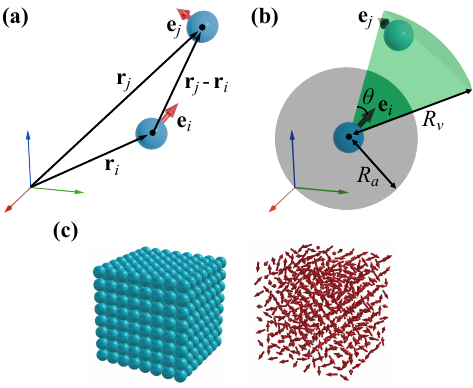}
    \caption{(a) A schematic illustration of  two spherical intelligent active Brownain particles (iABPs), labeled $i$ and $j$,  positioned at ${\bf r}_i$ and ${\bf r}_j$, with arrows indicating their  self-propulsion directions, $\mathbf{e}_i$ and $\mathbf{e}_j$, respectively. (b) A schematic  of the visual cone (green) with a range of $R_v$ and a vision angle $\theta$, along with the polar-alignment sphere (grey), which has a radius of $R_a$. (c) Initial configuration for the spherical iABPs simulation, where arrows in the inset represent the distribution of self-propulsion directions.}
    \label{fig:iabpsphere}
\end{figure}

We consider  a three-dimensional system of $N$ intelligent active Brownian spheres at positions ${\bf r}_i(t)$ and orientations ${\bf e}_i(t)$ at time $t$, with $i=1,\cdots, N$. We employ the overdamped Langevin equation to describe the translational motion of each particle 
\begin{equation}
 \dot{\mathbf{r}}_i =  v_0 \mathbf{e}_i + \beta D_t \mathbf{F}^{WCA}_i + \boldsymbol{\Gamma}_i,
 \label{eq:transmotion}
\end{equation}
where  $D_t$ represents the translational diffusion  coefficient, $\beta=1/k_BT$ denotes the inverse temperature with $k_B$ Boltzmann's constant and $T$ the temperature, and $v_0$ is a constant self-propulsion speed along the unit vector $\mathbf{e}_i(t)$ denoting the orientation of the particle. The force \( \mathbf{F}^{WCA}_i\) is given by \(-\sum_{j \neq i} \partial U_{WCA}(r_{ij})/\partial \mathbf{r}_{ij} \), where $U_{WCA}$ is the Weeks-Chandler-Andersen (WCA) potential defined as 
\begin{equation}
U_{WCA}(r_{ij}) = \begin{cases} 
4\epsilon \left[\left(\frac{\sigma}{r_{ij}}\right)^{12} - \left(\frac{\sigma}{r_{ij}}\right)^6\right] + \epsilon & r_{ij} \leq 2^{1/6}\sigma \\
0 & \text{otherwise,}
\end{cases}
\end{equation}
with $\mathbf{r}_{ij}=\mathbf{r}_i-\mathbf{r}_j$,  $r_{ij}=|\mathbf{r}_{ij}|$, $\epsilon$ representing the  interaction strength, and $\sigma$ the diameter of a particle. This potential describes the excluded-volume interactions between the self-propelled particles.

The term $\boldsymbol{\Gamma}_i$ represents a stochastic force acting on each particle, characterized by a zero mean $\langle \Gamma_{i,\alpha }(t)\rangle=0$ and second moment $\langle \Gamma_{i,\alpha }(t) \Gamma_{j,\beta }(t') \rangle = 2 D_t  \delta_{ij} \delta_{\alpha\beta} \delta(t - t')$, where  $\alpha$ and $\beta$ denote Cartesian coordinates. 

The rotational motion of each particle is described by
\cite{negi2024collective}
\begin{equation}
\dot{\mathbf{e}}_i = \mathbf{M}^v_i + \mathbf{M}^a_i + \boldsymbol{\Lambda}_{i}\times \mathbf{e}_i,
\label{eq:rotmotion}
\end{equation}
where $\dot{\mathbf{e}}_i(t)$ represents the time derivative of the self-propulsion direction, while $\mathbf{M}^v_i$ and $\mathbf{M}^a_i$ denote the vision-based self-steering and velocity-alignment torque, respectively, both of which will be explained in detail later. The stochastic term $\boldsymbol{\Lambda}_{i}$ satisfies the conditions $\langle \Lambda_{i,\alpha }(t)\rangle=0$ and $\langle \Lambda_{i,\alpha }(t) \Lambda_{j,\beta }(t') \rangle = 2 D_r \delta_{ij} \delta_{\alpha\beta} \delta(t - t')$, where $\alpha$ and $\beta$ refer to directions in spherical coordinates, $\mathbf{e}_\phi$  and  $\mathbf{e}_\theta$, and $D_r$ represents the rotational diffusion coefficient.

Following Ref. \citenum{negi2024collective}, the vision-based self-steering is described by 
\begin{equation}
\label{eq:sample2}
\mathbf{M}^v_i = \frac{\Omega_v}{N_{c,i}} \sum_{j \in VC} e^{-r_{ij}/R_0} \mathbf{e}_i \times (\mathbf{u}_{ij} \times \mathbf{e}_i),
\end{equation}
where $\Omega_v$ denotes a constant that signifies the strength of the vision-based self-steering, $N_{c,i}=\sum_{j \in VC} e^{-r_{ij}/R_0}$ represents a normalization constant, and the summation runs over all particles $j$ within the visual cone (VC). The visual cone is characterized by a range $R_v$ and a vision angle $\theta$, which sets the criterion for determining whether particle $j$ lies within this cone by $r_{ij} \leq R_v$ and $\mathbf{u}_{ij} \cdot \mathbf{e}_i \geq \cos{\theta}$ with $\mathbf{u}_{ij}=(\mathbf{r}_j-\mathbf{r}_i)/|\mathbf{r}_j-\mathbf{r}_i|$. These variables are depicted in Fig.~\ref{fig:iabpsphere}. The vision-based self-steering adjusts a particle's self-propulsion direction in response to the positions of particles within its  visual cone. As particles enter this cone, the particle reorients its self-propulsion direction toward the center of mass of the observed particles. The exponential term assigns higher weight to particles closest to the particle of interest, with a decay range of $R_0=R_v/4$.

The velocity-alignment torque reads 
\begin{equation}
\label{eq:sample3}
\mathbf{M}^a_i = \frac{\Omega_a}{N_{a,i}} \sum_{j \in PA} \mathbf{e}_i \times (\mathbf{e}_j \times \mathbf{e}_i),
\end{equation}
where $\Omega_a$ represents the strength of the velocity-alignment torque, $N_{a,i}$ denotes the number of particles within the polar-alignment sphere (PA), and the summation runs over all particles $j$ within this sphere. The polar-alignment sphere has a radius of $R_a$, which gives the criteria $r_{ij} \leq R_a$. These variables are also depicted in Fig.~\ref{fig:iabpsphere}. The velocity-alignment torque describes how the self-propulsion direction adjusts in response to the directions of neighbouring particles within the polar-alignment sphere. When other particles enter this sphere, the particle reorients its self-propulsion direction toward the average direction of these neighboring particles.

\begin{figure*}[ht]
    \centering
    \includegraphics[width=17cm]{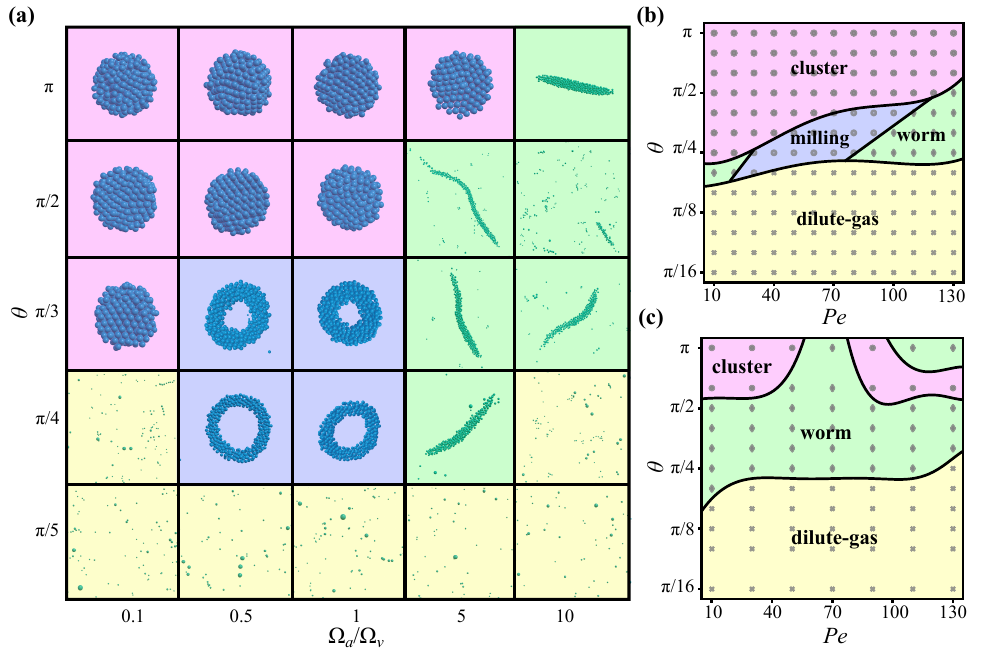} 
    \caption{(a) is the state diagram of intelligent active Brownian spheres in the relative strength of the velocity alignment and vision-based self-steering $\Omega_a/\Omega_v$ - vision angle $\theta$ plane at a  packing fraction $\Phi=6\times10^{-5}$ and Péclet number $\textrm{Pe}=v_0 \sigma/D_t=70$. The pink region denotes dense clusters, the green region depicts worm-like structures, the blue region corresponds to milling behavior, and the yellow region denotes dilute-gas structures. (b) and (c) are the state diagrams in the Péclet number $\textrm{Pe}$ - vision angle $\theta$ plane at same packing fraction and relative strength of the velocity alignment and vision-based self-steering (b) $\Omega_a/\Omega_v=1$, and (c) $\Omega_a/\Omega_v=5$.
}
    \label{fig:spherestatediagram} 
\end{figure*}

We employ Brownian dynamics simulations to explore the collective dynamics of intelligent active Brownian spheres. We numerically integrate the equations of motion in Eq.~\ref{eq:transmotion} and Eq.~\ref{eq:rotmotion}  with a  time step $\Delta t=10^{-5}\sigma^2/D_t$. We run the simulations for $3\times10^7$ steps to reach equilibrium. For specific parameter sets, additional steps are conducted.  The rotational diffusion coefficient is set to $D_r = 3D_t/\sigma^2$. We initially perform simulations of $N=512$ particles at a packing fraction of $\Phi=N V_p/V=6 \times 10^{-5}$, where $V_p=\pi\sigma^3/6$ denotes the volume of a single iABP, and $V$ represents the total volume of the simulation box.  The packing fraction is later increased  to examine the effect of varying box sizes. All simulations are conducted using periodic boundary conditions. Furthermore, we set the radius of the polar-alignment sphere to $R_a=2\sigma$, the characteristic decay range of the vision-based torque to $R_0=1.5 \sigma$, and the range of the visual cone to $R_v=6\sigma$. In our simulations, we fix the strength of the vision-based self-steering $\Omega_v=10^2 D_t/\sigma^2$, while adjusting  the relative strength of the velocity alignment $\Omega_a$ through the ratio $\Omega_a/\Omega_v$. We investigate the dynamic behavior of the system at varying  self-propulsion speed $v_0$, which is controlled through the Péclet number $\textrm{Pe}=v_0 \sigma/D_t$. The strength of the WCA potential is also adjusted with the Péclet number, using the relation  $\epsilon/k_BT = 1 + \textrm{Pe}$. 
Additionally, we  vary the vision angle $\theta$   from $0$ to $\pi$.   The simulations start with particles arranged in an $8\times8\times8$ simple cubic lattice at the center of the box with a lattice spacing of $\sigma$. Each particle is initialized with a random self-propulsion direction to avoid directional bias, as illustrated in Fig. \ref{fig:iabpsphere}(c). 

\subsection{State diagram}

To investigate the effects of velocity-alignment torque and vision-based self-steering, we set the Péclet number to $\textrm{Pe}=70$ and map out the state diagram by varying both the relative strength of  velocity alignment to vision-based self-steering $\Omega_a/\Omega_v$, and  the vision angle $\theta$. The results are summarized in Fig.~\ref{fig:spherestatediagram}(a).  We identify four distinct regions in the state diagram within the $\theta$-$\Omega_a/\Omega_v$ plane. 
The pink region represents dense clusters where particles densely aggregate and orient toward the center, driven by a wide vision angle  $\theta$ that enhances aggregation, while  low $\Omega_a/\Omega_v$ reduces mobility. The green region depicts  worm-like structures where particles aggregate and move cohesively. This is due to high $\Omega_a/\Omega_v$ which ensures  aligned propulsion, and wide vision angle $\theta$, which favors aggregation. In the yellow region, a narrow vision angle $\theta$ results in dilute-gas structures where particles are  randomly distributed in terms of both their directions and positions. The blue region corresponds to milling structures, where particles form rotating, doughnut-like structures, similar to ant milling behavior \cite{schneirla1944unique}. Notably, the cluster  at $\Omega_a/\Omega_v=5$ and $\theta=\pi$ remains mobile while others tend to be stationary. 
This state diagram reveals that a wider vision angle $\theta$ enhances aggregation, while higher values of $\Omega_a/\Omega_v$ lead to alignment of particle directions, consistent with previous findings from 2D simulations \cite{negi2024collective}.

We now investigate the effects of self-propulsion speed $v_0$ and vision angle $\theta$. We set  $\Omega_a/\Omega_v=1$, and explore the state diagrams by varying $\textrm{Pe}$ and $\theta$, as shown in Fig.~\ref{fig:spherestatediagram}. We again observe four distinct regions  in Fig.~\ref{fig:spherestatediagram}(b). An increase in $\theta$ can lead to the formation of dense cluster structures, similar as before. Increasing $\textrm{Pe}$ narrows the cluster region, and transforms milling structures into worm-like structures. Near the boundary between milling and worm structures, the system oscillates between these two states over time. In Fig.~\ref{fig:spherestatediagram}(c), where we fix $\Omega_a/\Omega_v=5$, the milling behavior is absent, and the cluster region diminishes, with the worm region predominating. This suggests that a higher $\Omega_a/\Omega_v$ ratio promotes the formation of worms. Additionally, the clusters in Fig.~\ref{fig:spherestatediagram}(b) are mobile, while those  in Fig.~\ref{fig:spherestatediagram}(c) tend to be stationary.

\begin{figure*}[htbp]
    \centering
    \includegraphics[width=17cm]{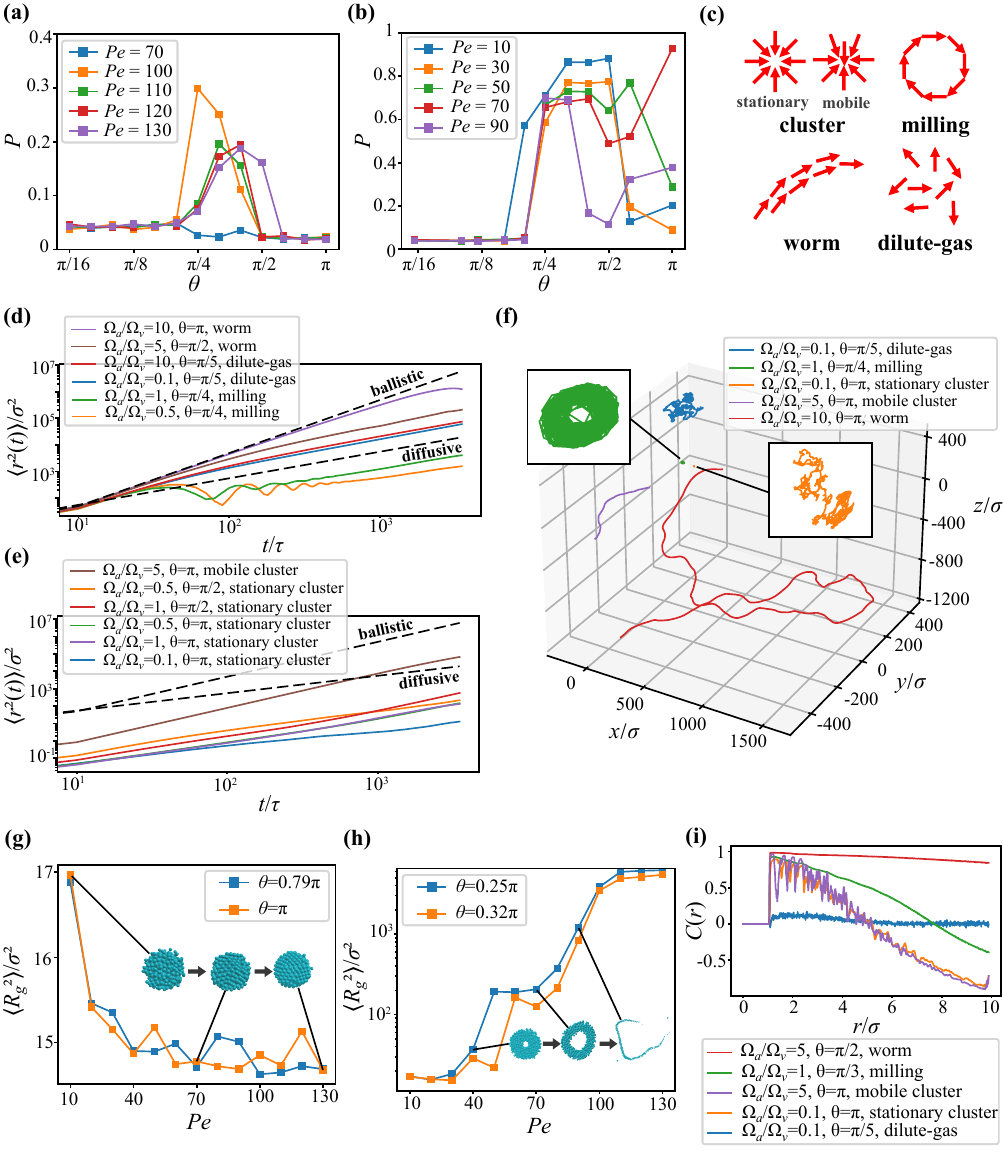} 
    \caption{(a) and (b) show the values of polarization $P$ as a function of $\theta$ for intelligent active Brownian spheres at a packing fraction of $\Phi = 6 \times 10^{-5}$ and relative strength of the velocity alignment and vision-based self-steering, (a) $\Omega_a/\Omega_v = 1$, and (b) $\Omega_a/\Omega_v = 5$, for varying Péclet number $\textrm{Pe}$ as indicated. (c) is a schematic illustration of the polarization of each particle, explaining the net polarization for different structures. (d) and (e) show the values of the mean square displacement $\langle r^2(t)\rangle$ for (d) worm-like, dilute-gas, milling structures, and (e) stationary and mobile dense clusters at the same packing fraction and Péclet number $\textrm{Pe} = v_0 \sigma/D_t = 70$, as a function of time $t/\tau$. These structures correspond to the ones displayed in Fig.~\ref{fig:spherestatediagram}(a). (f) shows the particle trajectories for various structures sampled over $3 \times 10^3\tau$. These structures also correspond to the ones displayed in Fig.~\ref{fig:spherestatediagram}(a). (i) shows the spatial angular correlation function $C(r)$ for these structures.
}
    \label{fig:sphereiabptest} 
\end{figure*}

\subsection{Polar order}
To analyze the orientation of the iABPs in the resulting structures, we measure the polarization $P$ defined as 
\begin{equation}
    P = \left\langle \frac{1}{N} \left|\sum_{i=1}^N \mathbf{e}_i \right|\right\rangle,
\end{equation}
where the summation runs over all particles, and the average is performed over time after equilibration. We plot the  polarization $P$ in Fig.~\ref{fig:sphereiabptest} as a function of vision angle $\theta$ for varying $\textrm{Pe}$ as labeled, corresponding to the systems displayed in the state diagrams in Figs.~\ref{fig:spherestatediagram}(b) and (c), respectively. 

Comparing Fig.~\ref{fig:sphereiabptest}(a) with Fig.~\ref{fig:spherestatediagram}(b), we observe that the dilute-gas, milling, and stationary cluster structures have polarization values of approximately $P\approx0$. In contrast,  worm-like structures exhibit significantly higher polarization values,    $P > 0$. Notably, these high $P$ values shift towards higher $\theta$ as $\textrm{Pe}$ increases, indicating that the region of worm-like structures expand into  higher $\theta$ regions while reducing the cluster region. When we compare Fig.~\ref{fig:sphereiabptest}(b) with Fig.~\ref{fig:spherestatediagram}(c), the dilute-gas region still shows $P \approx 0$, while worm-like structures have $P > 0$. Furthermore,  mobile clusters also exhibit, $P > 0$, distinguishing them from stationary clusters. 

In summary, the dilute-gas, stationary cluster and milling phases consistently yield  polarization values near zero, whereas the worm-like structures and mobile clusters exhibit  values larger than zero. Fig.~\ref{fig:sphereiabptest}(c) provides a schematic illustration of the polarization of each particle, explaining the net polarization observed for  different structures. Clusters display zero net polarization when all self-propulsion directions point towards the cluster center but show increased polarization when these directions deviate from the center. In the milling structures, the direction of each particle cancels out with that of the particle directly opposite it. In worm structures, particles tend to align in a specific direction. In dilute-gas structures, the random orientation of particles results in zero net polarization.  

\subsection{Mean square displacement}
We analyze the translational motion of the iABPs within each structure by measuring the mean square displacement (MSD)
\begin{equation}
\langle r^2(t)\rangle = \frac{1}{N} \sum_{i=1}^N \left \langle |\mathbf{r}_i(t + t_0) - \mathbf{r}_i(t_0)|^2 \right \rangle.
\end{equation}
Here, the summation runs over all particles, and the average is performed over different time windows. We present the MSD as a function of time for various structures of iABPs at a packing fraction $\Phi=6\times10^{-5}$ and Péclet number $\textrm{Pe}=v_0 \sigma/D_t=70$ in Fig.~\ref{fig:sphereiabptest}(d) and (e). These structures correspond to the ones displayed in Fig.~\ref{fig:spherestatediagram}(a). For comparison, we also plot the theoretical results for the MSD of a single ABP ~\cite{elgeti2015physics}, which behaves as  
\begin{equation}
\langle r^2(t) \rangle \approx \begin{cases} 
6D_tt+v_0^2t^2 &t \ll \tau_r \hspace{5mm} \text{(ballistic)}\\
6D_tt+v_0^2t/3D_r &t \gg \tau_r \hspace{5mm} \text{(diffusive)},
\end{cases}
\end{equation}
with $\tau_r=1/2D_r$  the rotational relaxation time. 
In Fig.~\ref{fig:sphereiabptest}(d), we present the MSD results for various structures, including worm-like, dilute-gas, and milling structures.  The worm-like structures, which occur  at high $\Omega_a/\Omega_v$, where alignment interactions dominate over vision-based self-steering, display nearly  ballistic motion. In contrast,  the dilute-gas structures exhibit  diffusive behavior at long times. Interestingly, milling structures show ballistic motion at short times, followed by oscillations at intermediate times due to their rotational milling behavior, before  becoming diffusive at longer times. In Fig.~\ref{fig:sphereiabptest}(e), we compare the MSDs of mobile and stationary dense clusters. Stationary clusters show very low MSD values, while mobile clusters exhibit significantly higher MSDs. The slow movement  of stationary clusters and the faster movement of  mobile clusters are also evident in  the particle trajectories shown in Fig.~\ref{fig:sphereiabptest}(f).  In stationary clusters and milling structures, particle trajectories  are localized, consistent with  their low MSD values. For  milling structures, the particles trace circular paths, with  each cycle resulting in approximately zero net displacement, explaining the periodic behavior seen in their MSDs. 
In contrast, the trajectory of the mobile cluster shows significant displacement over large timescales, while the worm-like structure exhibits long, relatively straight trajectories due to the highly aligned self-propulsion. The dilute-gas structures, on the other hand, display more  random movement, akin to Brownian motion.

\subsection{Radius of gyration}
The  size of the structure can be characterized by measuring the radius of gyration, defined by 
\begin{equation}
\langle R_g^2 \rangle = \frac{1}{N} \sum_{i=1}^N \langle\left( \mathbf{r}_i - \mathbf{r}_{cm} \right)^2\rangle,
\end{equation}
where the center of mass is given by $\mathbf{r}_{cm}=\frac{1}{N}\sum_{i=1}^N\mathbf{r}_i$, the summation runs over all particles, and the averaging is performed over time after equilibration. Fig.~\ref{fig:sphereiabptest}(g) shows $\langle R_g^2 \rangle$ evaluated for dense clusters observed at wide vision angles. We clearly observe that the cluster size  decreases as the Péclet number $\textrm{Pe}$ increases. This behavior can be attributed to the fact that lower $\textrm{Pe}$ values correspond to lower self-propulsion speeds, resulting in looser, less tightly bound clusters. Fig~\ref{fig:sphereiabptest}(f) demonstrates that the size of the milling structures observed for medium vision angles increases with increasing $\textrm{Pe}$. As $\textrm{Pe}$ increases, the boundary of the milling structure  becomes progressively thinner. Eventually, this thinning behavior destabilizes the structure, causing it to break apart into a worm-like structure.

\subsection{Spatial angular correlation function}

\begin{figure*}[ht]
    \centering
    \includegraphics[width=17cm]{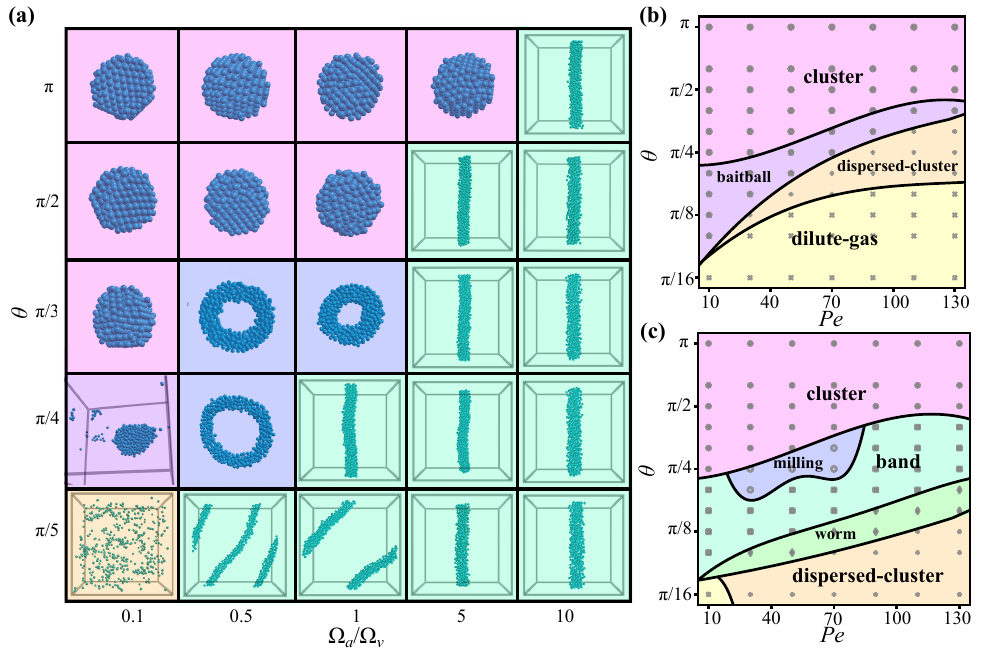} 
    \caption{(a) is the state diagram of intelligent active Brownian spheres in the relative strength of the velocity alignment and vision-based self-steering $\Omega_a/\Omega_v$ - vision angle $\theta$ plane at a  packing fraction $\Phi=6\times10^{-3}$ and Péclet number $\textrm{Pe}=v_0 \sigma/D_t=70$. The pink region denote dense clusters, the lilac region denotes baitball structures, the cyan region depicts band-like structures, the blue region corresponds to milling behavior, and the yellow region denotes dispersed clusters. (b) and (c) are the state diagrams in the Péclet number $\textrm{Pe}$ - vision angle $\theta$ plane at same packing fraction and relative strength of the velocity alignment and vision-based self-steering (b) $\Omega_a/\Omega_v=0.1$, and (c) $\Omega_a/\Omega_v=1$.
}
    \label{fig:spherestatediagramhighpk} 
\end{figure*}

To examine the spatial correlations of the self-propulsion directions, we measure the spatial angular correlation function~\cite{doliwa2000cooperativity} 
\begin{equation}
C(r) =\left\langle\frac{\sum_{i,j \neq i}^N \mathbf{e}_i \cdot \mathbf{e}_j \delta(r - |\mathbf{r}_i - \mathbf{r}_j)|}{\sum_{i,j \neq i} \delta(r - |\mathbf{r}_i - \mathbf{r}_j)|}\right\rangle,
\end{equation}
where the average is taken over time after equilibrium is reached. From Fig.~\ref{fig:sphereiabptest}(i), it is clear that the self-propulsion directions within worm-like structures are strongly correlated, indicating that particles tend to move in alignment with one another. The trends observed for milling and dense clusters show an  alignment of particles at close distances, followed by a decline into negative values, suggesting that particles on opposite sides of an aggregate have opposite self-propulsion directions. In dilute-gas structures, the alignment tends to be zero, reflecting a lack of angular correlation between particles.

\subsection{The effect of packing fraction}

In this section, we examine the effect of packing fraction $\Phi$  on the collective dynamics of the iABPs. We use the same parameter set as in Fig.~\ref{fig:spherestatediagram}, but increase the packing fraction from $\Phi=6 \times 10^{-5}$ to $6 \times 10^{-3}$. We plot the resulting state diagram in Fig.~\ref{fig:spherestatediagramhighpk}(a).
Notably, the dilute-gas structures disappear. This occurs  because at higher packing fractions, particles  find  and aggregate with each other more easily. We observed that all worm-like structures, and some milling structures,  transition into band-like structures, which are also found in other 2D models \cite{barberis2016large, peruani2016active,negi2024controlling}. The small box size affects worm-like structures, as they are too long to move freely. They loop back and connect to the opposite side of the box due to the periodic boundary conditions, forming these bands. These bands fall into two categories: single bands and multiple bands. Interestingly, all single bands align parallel to the box boundary, connecting across the shortest distance. Additionally, for the data set with $\Omega_a/\Omega_v = 0.1$ and $\theta = \pi/5$, dispersed clusters occur, where particles occasionally aggregate and then spread out. For the data set $\Omega_a/\Omega_v = 0.1$ and $\theta = \pi/4$, a baitball structure forms, where particles organize into a flexible, tornado-like shape, resembling  baitball behavior observed in fish \cite{lopez2012behavioural}.

We also investigate the effect of Péclet number $\textrm{Pe}$ and vision angle $\theta$ on the collective behavior of iABPs at a packing fraction  $\Phi=6\times10^{-3}$. We set the relative strength of the velocity-alignment and vision-based self-steering to $\Omega_a/\Omega_v=0.1$ where vision-based self-steering becomes the dominant interaction. We  present the resulting state diagram  in Fig.\ref{fig:spherestatediagramhighpk}(b). We observe that both the baitball and dispersed-cluster states occur between the dense cluster and dilute-gas states. This suggests that baitball and dispersed-cluster states are intermediate structures. When the velocity-alignment interaction strength is increased to $\Omega_a/\Omega_v$ to 1, as shown in Fig.~\ref{fig:spherestatediagramhighpk}(c), the baitball states and most of the dilute-gas regions vanish, giving rise to new structures such as milling, band-like, and worm-like structures.

\section{Intelligent active Brownian rods}
\label{sec:rods}
\subsection{Model system}

\begin{figure}[ht]
    \centering
    \includegraphics[width=8.4cm]{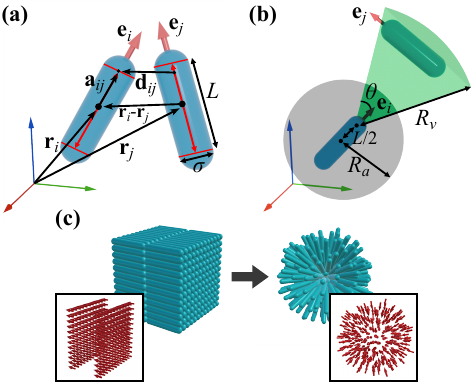}
    \caption{(a) A schematic illustration of two intelligent active Brownian rods, labeled $i$ and $j$,  located at their respective center-of-mass positions ${\bf r}_i$ and ${\bf r}_j$. The red lines represent their line segments. The vector $\mathbf{d}_{ij}$ indicates the shortest distance from particle $j$ to $i$, while the lever arm $\mathbf{a}_{ij}$ extends from the center of mass to the point of force application. The arrows along each particle's long-axis  denote their respective self-propulsion directions $\mathbf{e}_i$ and $\mathbf{e}_j$. Here, $L$ represents the length of the cylindrical part and $\sigma$ denotes the diameter of the hemispherical cap.  (b) A schematic of the visual cone (green) with a range of $R_v$ and a vision angle $\theta$, along with the polar-alignment sphere (grey), which has a radius of $R_a$.   The visual cone starts at the head of the particle, located at a distance of $L/2$ from the center of mass along its long-axis. (c) shows the procedures for generating the initial configurations for rod-like iABPs.}
    \label{fig:iabprod}
\end{figure}

Since some animals, such as fish and dogs, have anisotropic, rigid bodies, it is interesting to investigate how shape affects  the collective behavior of iABPs. To this end, we develop a model for rod-like iABPs, where the rods interact via  a shifted and truncated Kihara potential \cite{cuetos2002monte}
\begin{equation}
U_{Kih}(d_{ij}) = \begin{cases} 
4\epsilon \left[\left(\frac{\sigma}{d_{ij}}\right)^{12} - \left(\frac{\sigma}{d_{ij}}\right)^6\right] + \epsilon & d_{ij} \leq 2^{1/6}\sigma \\
0 & \text{otherwise,}
\end{cases}
\end{equation}
where $d_{ij}=|\mathbf{d}_{ij}|$ is the shortest distance between rod-like particle $i$ and $j$, with  $\mathbf{d}_{ij}$ representing the  shortest distance vector from $j$ to $i$, calculated using the  algorithm described in Ref. \citenum{vega1994fast}. 
In this model, two rod-like particles are treated as two line segments oriented along their long axes, intersecting at their center of mass. Each rod has a cylindrical part of length $L$  and feature a hemispherical cap with a diameter $\sigma$.  The self-propulsion direction for these rods is along their long axis, as illustrated in  Fig.~\ref{fig:iabprod}(a).

The translational motion of a rod-like intelligent active Brownian particle is described by the equation of motion 
\begin{equation}
 \dot{\mathbf{r}}_i =  v_0 \mathbf{e}_i + \beta \mathbf{D}_t \mathbf{F}^{Kih}_i + \boldsymbol{\Gamma}_i,
 \label{eq:transmotion2}
\end{equation}
where $\mathbf{D}_t$ denotes the diffusion tensor, the total force on particle $i$ due to the interactions between the rods reads $\mathbf{F}^{Kih}_i=\sum_{j \neq i} \mathbf{F}^{Kih}_{ij}$ with $\mathbf{F}^{Kih}_{ij}=-\partial U_{Kih}(d_{ij})/\partial \mathbf{d}_{ij}$ denoting the force exerted by rod $j$ on rod $i$, and the term $\boldsymbol{\Gamma}_i$ represents a stochastic force acting on particle $i$ with zero mean $\langle \Gamma_{i,\alpha }(t)\rangle=0$ and unit variance $\langle \Gamma_{i,\alpha }(t) \Gamma_{j,\beta }(t') \rangle = 2 \mathbf{D}_t  \delta_{ij} \delta_{\alpha\beta} \delta(t - t')$ with  $\alpha$ and $\beta$ representing Cartesian coordinates. 
Furthermore, the equation of motion for the rotational motion of particle $i$ reads 
\begin{equation}
\dot{\mathbf{e}}_i = \beta \mathbf{D}_r \mathbf{T}^{Kih}_i\times \mathbf{e}_i + \mathbf{M}^v_i + \mathbf{M}^a_i + \boldsymbol{\Lambda}_{i}\times \mathbf{e}_i,
\label{eq:rotmotion2}
\end{equation}
where $\mathbf{D}_r$ represents the rotational diffusion tensor,  and $\boldsymbol{\Lambda}_{i}$ represents a stochastic torque, satisfying the conditions $\langle \Lambda_{i,\alpha }(t)\rangle=0$ and $\langle \Lambda_{i,\alpha }(t) \Lambda_{j,\beta }(t') \rangle = 2 \mathbf{D}_r \delta_{ij} \delta_{\alpha\beta} \delta(t - t')$ with $\alpha$ and $\beta$ referring to directions in spherical coordinates, $\mathbf{e}_\phi$  and  $\mathbf{e}_\theta$ .
The torque is given by $\mathbf{T}^{Kih}_i = \sum_{j \neq i} \mathbf{F}^{Kih}_{ij} \times \mathbf{a}_{ij}$, where   $\mathbf{a}_{ij}$ is the lever arm from the center of mass of particle $i$ to the point of force application, as illustrated in Fig.~\ref{fig:iabprod}(a). 
Furthermore, we represent the translational and rotational diffusion tensors as 
\begin{equation}
\begin{aligned}
\mathbf{D}_t &= D_{\parallel} \mathbf{e}_i \mathbf{e}_i + D_{\perp}(\mathbf{I}- \mathbf{e}_i \mathbf{e}_i ) \\
\mathbf{D}_r &= D_r \mathbf{I}
\end{aligned}
\end{equation}
where $D_{\parallel}$ and $D_{\perp}$ are the translational diffusion coefficients parallel and perpendicular to the long-axis or self-propulsion direction $\mathbf{e}_i$ of the particle $i$. 
The diffusion coefficients are obtained from Refs.  \citenum{perrin1934mouvement, patti2012brownian}, and read 
\begin{equation}
\begin{aligned}
D_{\perp} &= \frac{D_t\left(2L^2 - 3\sigma^2\right)S + 2L \sigma}{16\pi\left(L^2 - \sigma^2\right)} \\
D_{\parallel} &= \frac{D_t\left(2L^2 - \sigma^2\right)S - 2L \sigma}{8\pi\left(L^2 - \sigma^2\right)} \\
D_r &= \frac{3 D_t\left(2L^2 - \sigma^2\right)S - 2L \sigma}{4\pi\left(L^4 - \sigma^4\right)}
\end{aligned}
\end{equation}

with
\begin{equation}
S = \frac{4}{\sqrt{L^2 - \sigma^2}} \ln \left( \frac{L + \sqrt{L^2 - \sigma^2}}{\sigma} \right).
\end{equation}

All other variables and constants are consistent with those defined in the previous section on spherical intelligent active Brownian particles.

\begin{figure*}[ht]
    \centering
    \includegraphics[width=17cm]{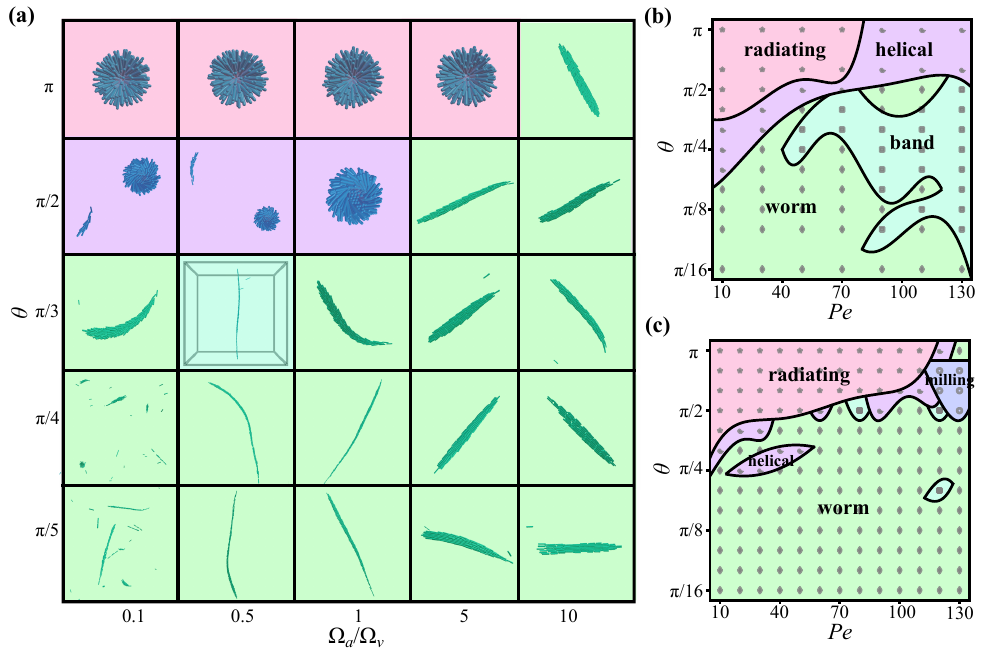} 
    \caption{(a) is the state diagram of intelligent active Brownian rods in the relative strength of the velocity alignment and vision-based self-steering $\Omega_a/\Omega_v$ - vision angle $\theta$ plane at a  packing fraction $\Phi=6\times10^{-5}$ and Péclet number $\textrm{Pe}=v_0 \sigma/D_t=70$. The pink region denotes radiating structures with the rods oriented towards the center, the green region depicts worm-like structures, the blue region corresponds to band-like structures,  and the purple region represents helical structures. (b) and (c) are the state diagrams in the Péclet number $\textrm{Pe}$ - vision angle $\theta$ plane at same packing fraction and relative strength of the velocity alignment and vision-based self-steering (b) $\Omega_a/\Omega_v=0.1$, and (c) $\Omega_a/\Omega_v=1$.
}
    \label{fig:rodstatediagram} 
\end{figure*}

We investigate the collective behavior of a system of intelligent active Brownian rods with a length-to-diameter ratio $L/\sigma = 5$ using Brownian dynamics simulations. We numerically integrate the equations of motion, as outlined in Eqs.~\ref{eq:transmotion2} and ~\ref{eq:rotmotion2},  with a  time step $\Delta t=10^{-5}\sigma^2/D_t$ for $3\times10^7$ steps to achieve equilibrium.  Simulations are conducted with $N=288$ particles at a packing fraction of $\Phi=N V_p/V=6 \times 10^{-5}$, where $V_p= \pi L\sigma^2/4 + \pi\sigma^3/6$ denotes the volume of a single rod, modeled as a spherocylinder.  We apply  periodic boundary conditions, and set the radius of the polar-alignment sphere to $R_a=4\sigma$, the characteristic decay range of the vision-based torque to $R_0=3 \sigma$, and the range of the visual cone to $R_v=12\sigma$. In all our simulations, we fix the strength of the vision-based self-steering $\Omega_v=10^2 D_t/\sigma^2$, while adjusting  the relative strength of the velocity alignment $\Omega_a$ through the ratio $\Omega_a/\Omega_v$.
 We investigate the dynamic behavior of the system at varying   Péclet number $\textrm{Pe}=v_0 \sigma/D_t$. The strength of the WCA potential is  adjusted with the Péclet number, using the relation  $\epsilon/k_BT = 1 + \textrm{Pe}$. 
Additionally, we  vary the vision angle $\theta$   from $0$ to $\pi$.   The simulations begin with all rods aligned at the center of the box, arranged in a  $12 \times 12 \times 2$ structure, with their self-propulsion directions uniformly oriented.  The simulation is initiated with $\textrm{Pe}=10$, $\Omega_a/\Omega_v=1$, $\theta=2\pi$,  resulting in a radiating structure. This structure serves as the initial configuration for all subsequent simulations, as illustrated in Fig.~\ref{fig:iabprod}(c). 

\subsection{State diagram}
We  first investigate the effects of velocity-alignment torque and vision-based self-steering on the collective behavior of intelligent active Brownian rods. We  set the Péclet number to $\textrm{Pe}=70$ and map out the state diagram as a function of the relative strength of  velocity alignment to vision-based self-steering $\Omega_a/\Omega_v$ and  the vision angle $\theta$. We present the state diagram in Fig.~\ref{fig:rodstatediagram}(a).  We identify four distinct regions in the state diagram within the $\Omega_a/\Omega_v-\theta$ plane. 
The pink region represents radiating structures where rods densely aggregate  and are all oriented towards the center. The formation of these radiating clusters is driven by a strong vision-based self-steering interaction at relatively low $\Omega_a/\Omega_v$ and a wide vision angle $\theta$.  The green region depicts  worm-like structures where rod-like particles aggregate and move cohesively. We find that the worm-like region is significantly larger compared to that of spherical iABPs, which can be attributed to the elongated body of the rod, making them more prone to alignment with neighboring particles. 
We also observe  a band-like structure at $\Omega_a/\Omega_v=0.5$ and $\theta=\pi/3$.
 The purple region represents helical structures, which differ from the radiating clusters due to their chirality. This behavior resembles the foraging patterns observed in dogs \cite{spin_dogs_2021}.

We now study the effects of Péclet number $\textrm{Pe}$ and vision angle $\theta$. We fix  $\Omega_a/\Omega_v$, and explore the state diagram by varying $\textrm{Pe}$ along with  $\theta$. 
The resulting state diagram is shown in Fig.\ref{fig:rodstatediagram}(b) for $\Omega_a/\Omega_v = 0.1$. We again observe four distinct regions, including radiating, helical, worm-like, and band-like structures. 
However,  when $\Omega_a/\Omega_v$ is increased to 1, as shown in Fig.~\ref{fig:rodstatediagram}(c), most of the helical and band-like regions vanish. These structures only appear sporadically near the boundaries between  radiating and worm-like regions. This observation suggests that these structures predominantly form at lower values of $\Omega_a/\Omega_v$. Interestingly, at higher Péclet numbers $\textrm{Pe}$, the system exhibits the emergence of  a milling structure.

\begin{figure*}[htbp]
    \centering
    \includegraphics[width=17cm]{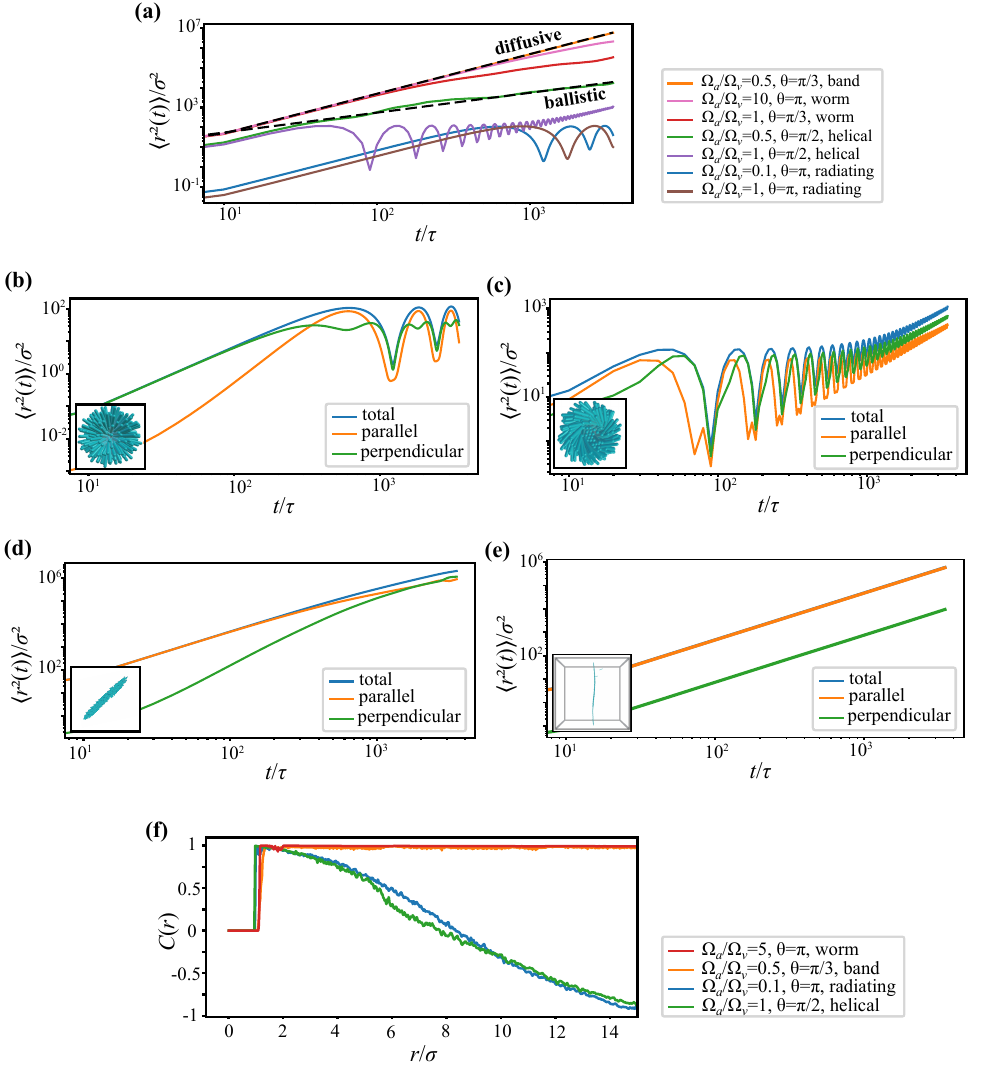} 
    \caption{ (a) shows the mean square displacement $\langle r^2(t)\rangle/\sigma^2$ for various structures of rod-like intelligent active Brownian particles at a packing fraction of $\Phi = 6 \times 10^{-5}$ and Péclet number $\textrm{Pe} = v_0 \sigma/D_t = 70$, as a function of time $t/\tau$. These structures correspond to the ones displayed in Fig.~\ref{fig:rodstatediagram}(a). (b)-(d) show the total mean square displacement $\langle r^2(t)\rangle/\sigma^2$, and its parallel and perpendicular components, for (b) radiating ($\Omega_a/\Omega_v = 0.1$, $\theta = \pi$), (c) helical ($\Omega_a/\Omega_v = 1$, $\theta = \pi/2$), (d) worm-like ($\Omega_a/\Omega_v = 10$, $\theta = \pi$), and (e) band-like ($\Omega_a/\Omega_v = 0.5$, $\theta = \pi/3$) structures at the same packing fraction $\Phi = 6 \times 10^{-5}$ and Péclet number $\textrm{Pe} = v_0 \sigma/D_t = 70$, as a function of time $t/\tau$. These structures correspond to the ones displayed in Fig.~\ref{fig:rodstatediagram}(a). (f) shows the spatial angular correlation function $C(r)$ for these structures.
}
    \label{fig:rodiabptest} 
\end{figure*}

\subsection{Mean square displacement}
We measure the MSD to characterize the translational motion of rod-like iABPs.  We present the MSD as a function of time for various structures of iABPs, such as band-like, worm-like, helical, and radiating structures, at a packing fraction $\Phi=6\times10^{-5}$ and Péclet number $\textrm{Pe}=v_0 \sigma/D_t=70$ in Fig.~\ref{fig:rodiabptest}(a). These structures correspond to the ones displayed in Fig.~\ref{fig:rodstatediagram}(a), along with the theoretical predictions of a single ABP in the ballistic and diffusive regime \cite{elgeti2015physics}. The band-like structure, which occurs at relatively narrow vision angle, shows ballistic motion. In contrast, worm-like structures only display ballistic behavior for sufficiently high $\Omega_a/\Omega_v$ or narrow vision angle, but become diffusive otherwise.  
We observe that the MSD values are significantly smaller for helical structures and orders of magnitude smaller for radiating clusters, compared to band-like and worm-like structures. Interestingly, helical structures show ballistic motion at short times, followed by either diffusive motion or oscillatory behavior at intermediate times due to the rotational motion of the cluster, while the radiating structures all become oscillatory at long times. In addition, we can decompose the translational motion of rod-like iABPs into motion parallel and perpendicular to their long axis. In Fig.~\ref{fig:rodiabptest}(b)-(d), we plot the perpendicular and parallel components of the MSD for radiating, helical, and worm-like structures together with the total MSD for comparison. Fig.~\ref{fig:rodiabptest}(b)-(d) show that for radiating structures the total MSD at short timescales (less than $100\tau$) is primarily driven by perpendicular motion, as the rods exhibit only small wiggling motions around a fixed point. At longer time scales (larger than $1000\tau$), both  perpendicular and parallel motions contribute to the MSD as the entire structure becomes mobile. In contrast, helical structures display a more balanced contribution  from both types of motion initially, with perpendicular motion becoming increasingly dominant over  time as the  structure tends to drift in a perpendicular direction. Worm-like structures are initially dominated by  parallel motion due to ballistic movements, but over time, perpendicular motion takes over  as the worm-like structure begins to  turn. Finally, for straight-moving, band-forming structures, parallel motion remains the primary contributor to the total MSD throughout the entire simulation.

\subsection{Spatial angular correlation function}
Finally, we also evaluate the spatial angular correlation function of the rod-like iABPs, and  present the results in Fig.~\ref{fig:rodiabptest}(f). The analysis reveals strong alignment of self-propulsion directions within both band-like and worm-like structures, indicating that the particles move in coordination. In contrast, for radiating and helical structures, the correlation is high at short distances but drops rapidly to negative values as the distance increases. This behavior resembles the trends observed for spherical particles, where particles on opposite sides of a cluster exhibit opposing self-propulsion directions, resulting in the rapid decline in correlation. 

\section{Intelligent active Brownian worms}
\label{sec:worms}
\subsection{Model system}

\begin{figure}[ht]
    \centering
    \includegraphics[width=8.4cm]{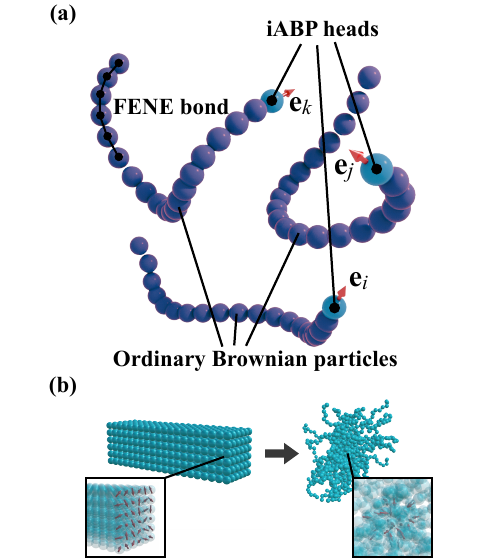}
    \caption{(a) Schematic picture of three intelligent active Brownian worms. Each worm features an iABP head and consists of particles connected by the FENE potential. The red arrows, $\mathbf{e}_i$, $\mathbf{e}_j$ and $\mathbf{e}_k$, denote the self-propulsion direction of the iABP heads of worm $i$, $j$, and $k$, respectively. (b) shows the procedures for generating the initial configurations for iABP worms.}
    \label{fig:iabpworm}
\end{figure}

Inspired by animals such as snakes and worms, which possess long and flexible bodies, we propose a model for worm-like iABPs. The polymer chains are represented as Brownian particles connected via the finitely extensible non-linear elastic (FENE) potential \cite{kremer1990dynamics}
\begin{eqnarray}
  U_{FENE}(r_{ij}) = 
 \begin{cases} 
-0.5kL_M^2 \ln\left[1 - \left(\frac{r_{ij}}{L_M}\right)^2\right], & r_{ij} \leq L_M \\
\infty, & \text{otherwise,}
\end{cases}  
\end{eqnarray}
where $k$ is the spring constant that represents the strength of the interaction, and $L_M$ is the maximum bond length that limits the extension of the bond. The dynamics of these polymer chains are described by the equation of motion 
\begin{equation}
\dot{\mathbf{r}}_i = \beta D_t \mathbf{F}^{WCA}_i + \beta D_t \mathbf{F}^{FENE}_i+\boldsymbol{\Gamma}_i,
\end{equation}
where $\mathbf{F}^{FENE}_i=-\sum_{bond} \partial U_{FENE}(r_{ij})/\partial \mathbf{r}_{ij}$, with $\sum_{bond}$ denoting the sum over all particles connected to $i$. Next, spherical iABPs are attached to these polymer chains, forming iABP worms, as illustrated in Fig.~\ref{fig:iabpworm}(a). The equations of motion for these  iABP heads are given by 
\begin{equation}
\begin{split}
\dot{\mathbf{r}}_i &= v_0 \mathbf{e}_i + \beta D_t \mathbf{F}^{WCA}_i + \beta D_t \mathbf{F}^{FENE}_i+ \boldsymbol{\Gamma}_i \\
\dot{\mathbf{e}}_i &= \mathbf{M}^v_i + \mathbf{M}^a_i + \boldsymbol{\Lambda}_{i} \times \mathbf{e}_i.
\end{split}
\end{equation}

\begin{figure*}[ht]
    \centering
    \includegraphics[width=17cm]{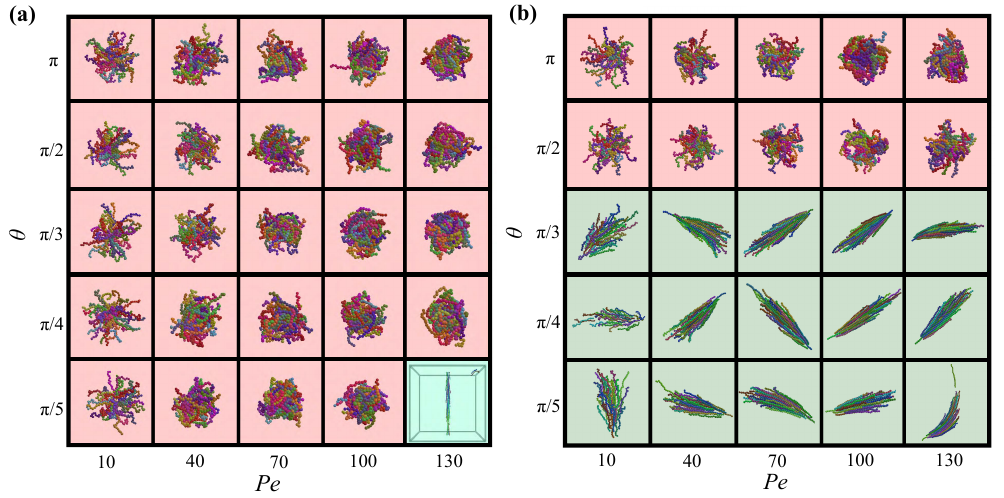} 
    \caption{The state diagram of intelligent active Brownian worms in the Péclet number $\textrm{Pe}$ - vision angle $\theta$ plane at a  packing fraction $\Phi=6\times10^{-4}$ and relative strength of the velocity alignment and vision-based self-steering (a) $\Omega_a/\Omega_v=0.1$, and (b) $\Omega_a/\Omega_v=1$.  The pink region denotes dense clusters (or worm blobs), the green region depicts band-like or streamlined structures. 
}
    \label{fig:wormstatediagram} 
\end{figure*}

In our model, we assume that the vision-induced torque $\mathbf{M}^v_i$ acting on  the iABP head, is  influenced by both iABP heads and the tails composed of  ordinary Brownian particles. Additionally, the worm has self-recognition capabilities, which prevent it from  pursuing its own tail when it enters the visual cone. Finally, the velocity-alignment torque $\mathbf{M}^v_i$ for the  iABP head is affected exclusively by other iABP heads. As a result, the summation in Eq.~(\ref{eq:sample2}) for $\mathbf{M}^v_i$  includes all particles within the visual cone except those belonging to    the same chain, while the summation in Eq.~(\ref{eq:sample3}) for $\mathbf{M}^a_i$ is restricted to  other iABP heads within the polar-alignment sphere. The definitions of  other variables and constants remain as described in the previous section.

We investigate the dynamics of 36 worms, each consisting of one iABP head and 19 ordinary Brownian particles. We set the packing fraction to $\Phi=6 \times 10^{-4}$. 
Given that the worms are much larger than a single iABP, we set  the radius of the polar-alignment
sphere to $R_a= 5 \sigma$, the characteristic decay range of the vision-based torque to $R_0=4\sigma$,  and the range of the visual cone to $R_v= 16\sigma$. We apply periodic boundary conditions. The maximum bond length is set to  $L_M=1.5\sigma$, and the spring constant is $k=30\epsilon/\sigma^2$, ensuring tight connections between the particles. Initially, all worms are arranged in the center of the box, aligned to form a $6 \times 6 \times 20$ structure with $\sigma$ spacing between them. The simulation is  started at a Péclet number $\textrm{Pe}=10$, relative strength of the velocity alignment and  vision-based self-steering $\Omega_a/\Omega_v=0.1$, vision angle $\theta=2\pi$, for $2\times10^7$ steps, using a time step of $\Delta t=5\times10^{-6}\sigma^2/D_t$. The resulting inverted micelle structure is then employed as the initial configuration for all subsequent simulations, as illustrated in Fig.\ref{fig:iabpworm}(b). 

\subsection{State diagram}

We first examine the collective behavior of worm-like iABPs at fixed $\Omega_a/\Omega_v$. The state diagrams  for worm-like iABPs as a function of Péclet number $\textrm{Pe}$ and vision angle $\theta$  are presented in Fig.~\ref{fig:wormstatediagram}. In the state diagram at $\Omega_a/\Omega_v=0.1$, the predominant structures are worm blobs, where the  worms aggregate in various structures. At very low  $\textrm{Pe}$, their heads orient toward the center while their tails wiggle around. In other cases, the entire bodies aggregate densely. Interestingly, the clusters resemble micellar structures and remain finite because of the steric hindrance caused by the tails. A band-like structure emerges at $\textrm{Pe}=170$ and $\theta=\pi/5$. As the ratio $\Omega_a/\Omega_v$ increases to 1, we begin to observe  streamlined structures, in which  all the worms aggregate and move in a specific direction.

\begin{figure*}[ht]
    \centering
    \includegraphics[width=17cm]{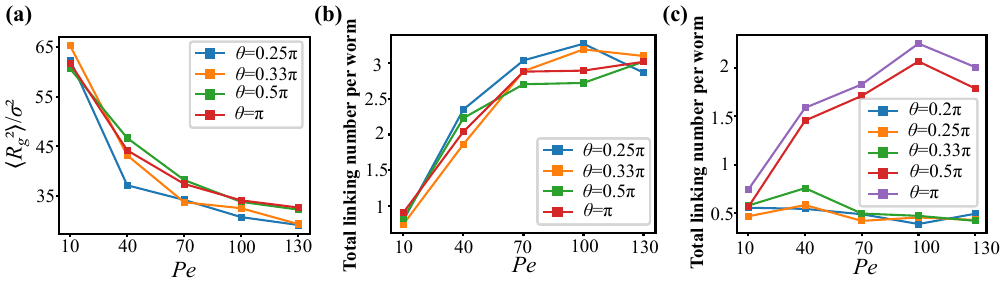} 
    \caption{(a) The radius of gyration $\langle R_g^2\rangle$  as a function of Péclet number $\textrm{Pe}$ for dense clusters or worm blobs of intelligent active Brownian worms corresponding to those presented in Fig.~\ref{fig:iabpworm}(a). (b) and (c) show linking number as a function of $\textrm{Pe}$ for varying vision angles $\theta$ as labeled, corresponding to (b) worm blobs at $\Omega_a/\Omega_v=0.1$ as shown in  Fig.~\ref{fig:iabpworm}(a), and (c) worm blobs and streamlined structures at $\Omega_a/\Omega_v=1$ depicted in Fig.~\ref{fig:iabpworm}(b).
}
    \label{fig:wormtest} 
\end{figure*}

\subsection{Radius of gyration}
We  characterize the size of the blobs by analyzing the radius of gyration $R_g$, as shown in Fig.~\ref{fig:wormtest}(a). Consistent with the findings for the clusters of spherical iABPs, the size of the blobs decreases with increasing  $\textrm{Pe}$, indicating that the blobs become more densely aggregated  at higher values of $\textrm{Pe}$.

\subsection{Linking number}

The phenomena observed in the radius of gyration $R_g$  can be explained by a metric known as the linking number, which measures the entanglement strength of polymers. The linking number for a pair of worms  can be calculated using the equation \cite{qu2021fast}
\begin{equation}
Lk_{abs} = \frac{1}{4\pi} \sum_{i,j} \left| \frac{\Delta \mathbf{r}_i \times \Delta \mathbf{r}_j \cdot (\mathbf{r}_i - \mathbf{r}_j)}{|\mathbf{r}_i - \mathbf{r}_j|^3} \right|,
\end{equation}

where $\mathbf{r}_i$ represents the  positions of the particles of a worm, and $\Delta \mathbf{r}_i$ denotes the vector $\mathbf{r}_{i+1} - \mathbf{r}_{i}$. We take the absolute value because we are primarily interested in  the entanglement strength. It is important to note that this equation is designed to calculate the linking number of two closed chains. To adapt it  for open chains, we have to exclude the last particle position to ensure that  the number of $\mathbf{r}_i$ matches that of  $\Delta \mathbf{r}_i$. Finally,  the total linking number can be calculated by summing up the pair linking numbers. We present the total linking number for worm blobs at a relative strength of the velocity alignment and vision-based self-steering $\Omega_a/\Omega_v=0.1$ in  Fig.~\ref{fig:wormtest}(b), indicating that higher $\textrm{Pe}$ corresponds to stronger entanglement, which explains the increased density of the blobs.  Fig.~\ref{fig:wormtest}(c) shows the linking number of worm blobs and streamlined structures, demonstrating that the linking number for streamlined structures remains consistently low, whereas the linking number for worm blobs increases with $\textrm{Pe}$.

\section{Conclusions}
\label{sec:conclusions}
In this study, we investigated the collective behaviors of intelligent active Brownian spheres, rods, and worms in three dimensions using Brownian dynamics  simulations. Our primary focus was on several  key parameters: the visual angle $\theta$, which determines the angle of the visual cone; the Péclet number $\textrm{Pe}$, which influences the self-propulsion speed; and the ratio $\Omega_a/\Omega_v$, which balances  velocity alignment and visual perception capabilities.

First, we investigated the collective behavior of spherical iABPs and observed the spontaneous formation of dense  clusters, worm-like structures, milling configurations, and dilute-gas structures. By measuring the total polarization $P$, we found that $P\approx 0$ for milling, dilute-gas, and stationary clusters, while $P>0$ for worm-like and mobile clusters. Measurements of the mean square displacement (MSD)  revealed that particles within worm-like structures exhibited nearly ballistic motion, whereas particles in dilute-gas structures  demonstrated motion  close to diffusive behavior. Milling structures exhibited periodic behavior, stationary clusters had very low MSD values, and mobile clusters gradually transitioned to diffusive motion. 
Measurements of the radius of gyration indicated  that clusters became denser while milling structures became thinner with increasing $\textrm{Pe}$. Analysis of the spatial angular correlation functions  indicated that the self-propulsion directions of worm-like structures were highly correlated, while dilute-gas structures exhibited  no correlations.  Milling and cluster structures demonstrated positive correlations that gradually shifted to  negative values. Additionally, at higher packing fraction, we observed new patterns, such as dispersed clusters, band-like structures, and baitball structures.

Furthermore, we transformed the spherical iABPs into rod-like shapes by introducing the Kihara potential. We observed worm-like, band-like, helical, and radiating structures. The motion of these structures could be decomposed into parallel and perpendicular components, each contributing differently to the overall dynamics. Measurements of the spatial angular correlation functions  showed that the self-propulsion directions of worm-like and band-like structures were highly correlated, while helical and radiating structures exhibited values that shifted from positive to negative.

Lastly, by connecting spherical iABPs with ordinary Brownian particles to create iABP worms, we observed blob-like, band-like, and streamlined structures. Measurements of the radius of gyration and linking number  revealed that the size of the blobs decreased with increasing Péclet number $\textrm{Pe}$ due to stronger entanglement, while the linking number for streamlined structures remained consistently low.

The observed structures closely resemble behaviors seen in animal groups, such as ants forming milling structures, fish forming baitballs, dogs exhibiting helical structures, and worms entangling into blobs. By aligning parameters such as the visual angle $\theta$ and Péclet number $\textrm{Pe}$ with those of real animals, these models have the  potential to predict collective animal behaviors. With advances in synthetic techniques, we can design various nano-robots inspired by these findings, equipping them with visual perception and velocity alignment capabilities. In addition, we can develop self-propelled colloidal particles of which their motility can be adjusted through an external feedback loop. Recent research indicates that multicellular systems can be understood through the framework of active matter.  Considering that multicellular systems originated from single organisms capable of sensing light or water, this research may also offer insights into the origins of multicellular systems.

\section*{Conflicts of interest}
There is no conflict of interest.

\section*{Data availability}
The data and calculations that support the study are available on reasonable requests. 

\section*{Acknowledgements}
M.D. acknowledges funding from the European Research Council (ERC) under the European Union’s Horizon 2020 research and innovation programme (Grant agreement No. ERC-2019-ADG 884902 SoftML). We also acknowledge financial support from NWO and from Canon Production Printing Netherlands B.V., FIP2 project KICH2.V4C.20.001. 



\balance


\bibliography{rsc} 
\bibliographystyle{rsc} 

\end{document}